\documentclass{mem}
\usepackage{natbib}
\usepackage{txfonts}
\usepackage{balance}
\usepackage{flushend}
\usepackage{graphicx}
\usepackage{tablefootnote}
\usepackage[breaklinks,pdftex]{hyperref}
\idline{75}{282}
\begin{document}
\def\teff{$T\rm_{eff }$}
\def\kms{$\mathrm {km s}^{-1}$}

\title{
LST-1 follow-up of the exceptionally bright gamma-ray burst GRB~221009A
}

\author{
A. \,Aguasca-Cabot\inst{1} \and A. Carosi\inst{2} \and A. Donini\inst{2} \and S. Inoue\inst{3} \and Y. Sato\inst{4} \and M. Seglar Arroyo\inst{5} \and K. Terauchi\inst{6} \and  P. Bordas\inst{1} \and M. Ribó\inst{1} for the CTAO-LST Project
}

\institute{
Departament de Física Quàntica i Astrofísica, Institut de Ciències del Cosmos, Universitat de Barcelona, IEEC-UB, Martí i Franquès, 1, 08028, Barcelona, Spain. \email{arnau.aguasca@fqa.ub.edu}
\and
INAF - Osservatorio Astronomico di Roma, Via di Frascati 33, 00040, Monteporzio Catone, Italy 
\and
Institute for Cosmic Ray Research, University of Tokyo, 5-1-5, Kashiwa-no-ha, Kashiwa, Chiba 277-8582, Japan
\and
Department of Physical Sciences, Aoyama Gakuin University, Fuchinobe, Sagamihara, Kanagawa, 252-5258, Japan
\and
Institut de Fisica d’Altes Energies (IFAE), The Barcelona Institute of Science and Technology, Campus UAB, 08193 Bellaterra (Barcelona), Spain
\\
The remaining affiliations can be found at the end of the paper.
}

\authorrunning{A. Aguasca-Cabot}
\titlerunning{LST-1 follow-up of the gamma-ray burst GRB~221009A}

\date{Received: XX-XX-XXXX (Day-Month-Year); Accepted: XX-XX-XXXX (Day-Month-Year)}

\abstract{
On 9 October 2022, the brightest gamma-ray burst (GRB) ever recorded (GRB~221009A) was detected. Its remarkably bright emission, partially due to its close distance to Earth ($z\sim0.15$), makes this GRB a unique event. The outstanding characteristics of GRB~221009A, including the TeV detection by the LHAASO experiment, triggered deep follow-up observations of the source across all wavebands, including very-high-energy gamma rays with the first Large-Sized Telescope (LST-1) of the future Cherenkov Telescope Array Observatory. LST-1 observations started about one day after the onset of the prompt emission, under strong moonlight conditions. This resulted in a hint of a signal with a statistical significance of about 4$\sigma$. The monitoring of this source continued until the end of November 2022. This constitutes the deepest observation campaign performed on a GRB with LST-1. Here we show the results of this follow-up campaign.
}
\maketitle{}

\section{Introduction}
GRB~221009A is an exceptionally long gamma-ray burst (GRB), the brightest GRB observed up to date. Estimated to be a nearby burst \citep[$z=0.1505$;][]{GTCGCN2022}, its bright luminosity and proximity make this GRB an out-of-ordinary event \citep{Williams2023swift}.

GRB~221009A was detected on 9 October 2022 \citep{Swift2022GCN_xraytrans,GBM2022GCN}. 
At very-high-energy gamma rays (VHE; $E>100\,$ GeV), it was first observed by water Cherenkov detectors (WCDs). The GRB location was serendipitously covered by the LHAASO field-of-view (FoV) when the burst occurred, being detected up to 13\,TeV with more than 60,000 photons \citep{LHAASO2022GCN, LHAASO2023WCDA, LHAASO2023KM2A}. Remarkably, it is the first GRB detected by a WCD and the first detection of the onset of the VHE afterglow emission. HAWC observed it about 8 hours later without yielding a significant detection of the event \citep{HAWC2022GCN}. Follow-up observations with imaging atmospheric Cherenkov telescopes (IACTs) were delayed due to the presence of the Moon, which affected the data-taking operations.

GRB~221009A was extensively followed up across the electromagnetic spectrum, offering valuable insights into the prompt and afterglow emission of this event, linked to jet physics in long GRBs. 
Observations suggest that the emission from GRB~221009A may be associated with a structured jet \citep[][]{Ren2024,Zhang2023arXiv,Zheng2024}. Further multiwavelength data are required to constrain the parameters of these sophisticated models.

The first Large-Sized Telescope (LST-1) of the upcoming Cherenkov Telescope Array Observatory (CTAO) conducted an extensive observation campaign on this GRB, starting data taking during the full moon period. In this work, we report the LST-1 results on this deep follow-up, which contains the first observations by an IACT on this event in a time period without published VHE gamma-ray data. In Sect.~\ref{sect:obs} the observation campaign with LST-1 is described. The details of the analysis are explained in Sect.~\ref{sect:ana} while results are presented in Sect.~\ref{sect:res}. Finally, we provide the conclusions in Sect.~\ref{sect:conc}.

\section{Observations with LST-1}
\label{sect:obs}

The LSTs are the largest telescope type of CTAO. They are ideally suited for transient GRB observations due to (a) their fast repositioning capabilities to slew at any pointing direction in the sky within 20\,s, (b) the low energy threshold allowing the detection of gamma rays down to 20\,GeV, and (c) the wide camera FoV of 4.5\,deg \citep{Abe:2023Vc}.

LST-1 observations on GRB~221009A provide early-time data on this GRB starting at $1.33\,$d after the burst trigger \citep[$T_0=2022$-10-09\,13:16:59.99 UTC;][]{Lesage_2023}, being the first follow-up observation by an IACT. Data were collected over several nights until the end of November 2022 under varying observing conditions. The initial two observation nights were notably affected by high night-sky-background (NSB) observing conditions, while subsequent observations with the Moon below the horizon were recorded on later nights until the end of October. Finally, some data with the presence of the Moon preceded the observations in November, which were taken without it. 
We report here the observations obtained in October 2022. In particular, we consider the observations for the first two days when the Moon was above the horizon, while the Moon was not present for the rest of the data (see Table~\ref{tab:obs_table}).

\begin{table*}[t!]
\caption{Observations of GRB~221009A reported in these proceedings. For each observation day, we show the date of the evening before the observations, the time delay between the start of data taking and the trigger time ($T_0$), the minimum and maximum zenith angle of the observations and the total observation time.}
\begin{center}
\begin{tabular}{lcccc}
\hline
Date (YYYY-MM-DD) & $T-T_0$ (d) & Min. zenith (deg) & Max. zenith (deg) & Observation time (h)\\
\hline
2022-10-10$^{a}$ & 1.33 & 31 & 54 & 1.75\\
2022-10-12$^{b}$ & 3.33 & 34 & 52 & 1.42\\
2022-10-15 & 6.30 & 25 & 52 & 0.80\\
2022-10-16 & 7.32 & 34 & 65 & 2.35\\
2022-10-17 & 8.30 & 28 & 60 & 2.41\\
2022-10-23 & 14.30 & 34 & 61 & 2.01\\
2022-10-25 & 16.33 & 45 & 59 & 1.18\\
2022-10-26 & 17.32 & 42 & 58 & 1.42\\
\hline
\end{tabular}
\end{center}
\footnotesize{$^a$ Observations under very high NSB observing conditions.}\\
\footnotesize{$^b$ Observations under high NSB observing conditions.}\\
\label{tab:obs_table}
\end{table*}

\section{Data processing and analysis}
\label{sect:ana}

We employ two different analysis approaches to ensure the best possible performance under the diverse observing conditions that affected the observation campaign. One is devoted to the analysis of data obtained in nominal observing conditions (dark night; Moon below the horizon), and another is appropriate for the high NSB conditions in the first two observation nights.

The observations in nominal conditions are analysed using the standard source-independent analysis \citep{Abe_2023_LSTPerformance}. For observations affected by moonlight, a tailored analysis is developed to diminish the influence of the increased spurious NSB triggers: (a) we calibrated the data accounting for fast-changing observing conditions, (b) optimised the signal extractor and image cleaning step and (c) performed an observation-by-observation analysis. One of the consequences of the high NSB conditions triggered by significant moonlight is the increase of the energy threshold of the analysis up to a few hundreds of GeV (about an order of magnitude higher than for the standard analysis).

We note that the analysis of the data obtained on GRB~221009A is performed using two independent analysis chains to cross-check and validate the results. Both of them were done using a 1-D analysis with three OFF regions to assess the signal and estimate the emission of GRB~221009A. We utilise the Gammapy package for this purpose \citep{gammapy_v1.0,gammapyAA2023}. In this work we show the results of one of the independent analysis chain. Further details on the analyses are provided in \citet{2025ApJ...988L..42A}.

\section{Preliminary results}
\label{sect:res}
\begin{figure}[t!]
\resizebox{\hsize}{!}{\includegraphics[clip=true]{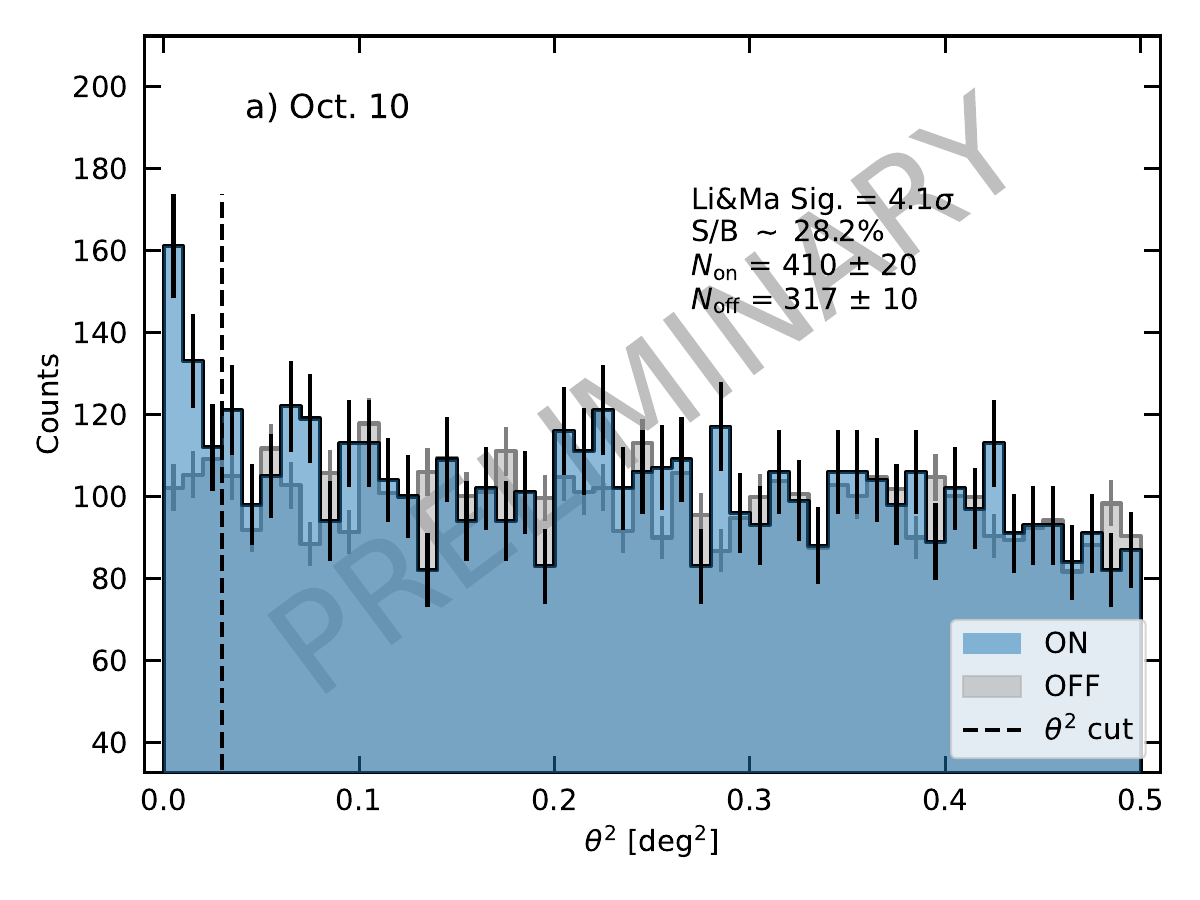}}\\
\resizebox{\hsize}{!}{\includegraphics[clip=true]{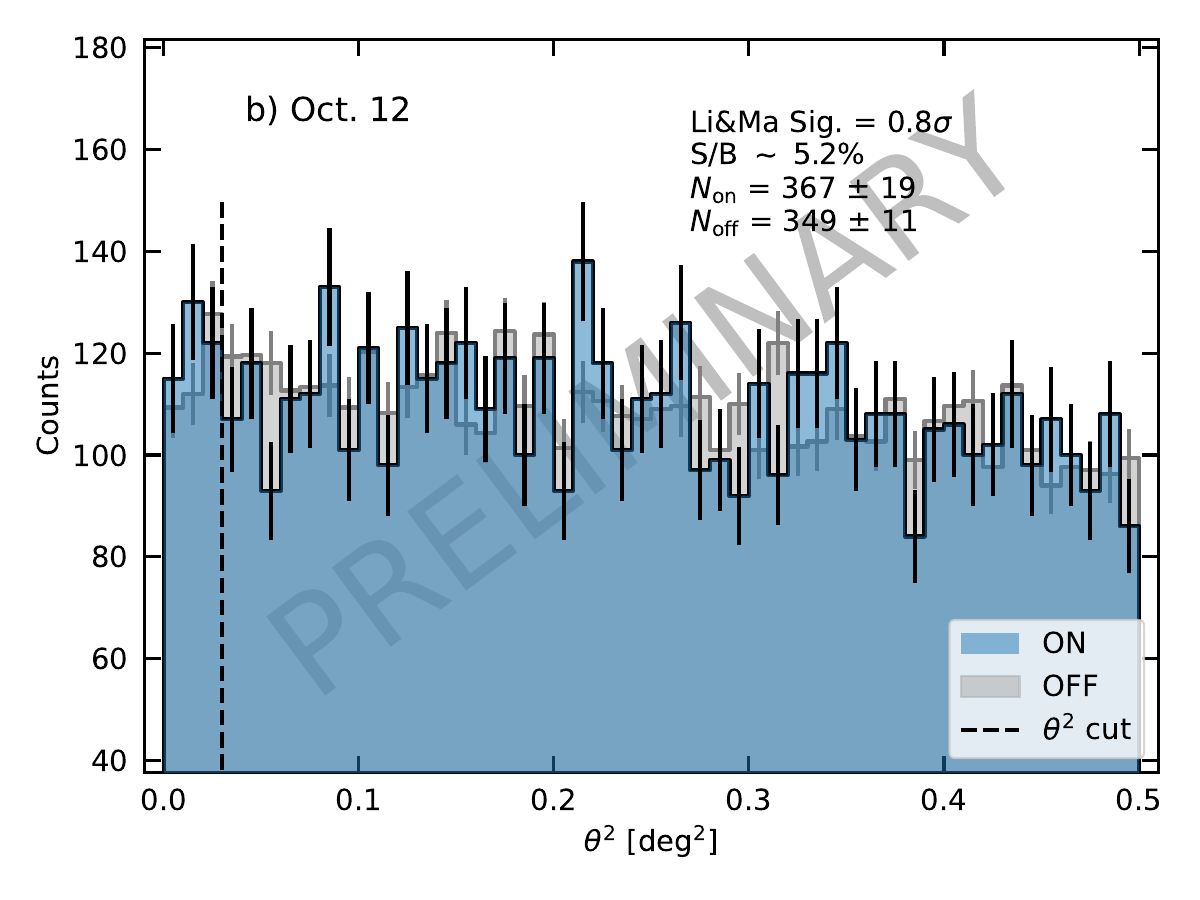}}\\
\resizebox{\hsize}{!}{\includegraphics[clip=true]{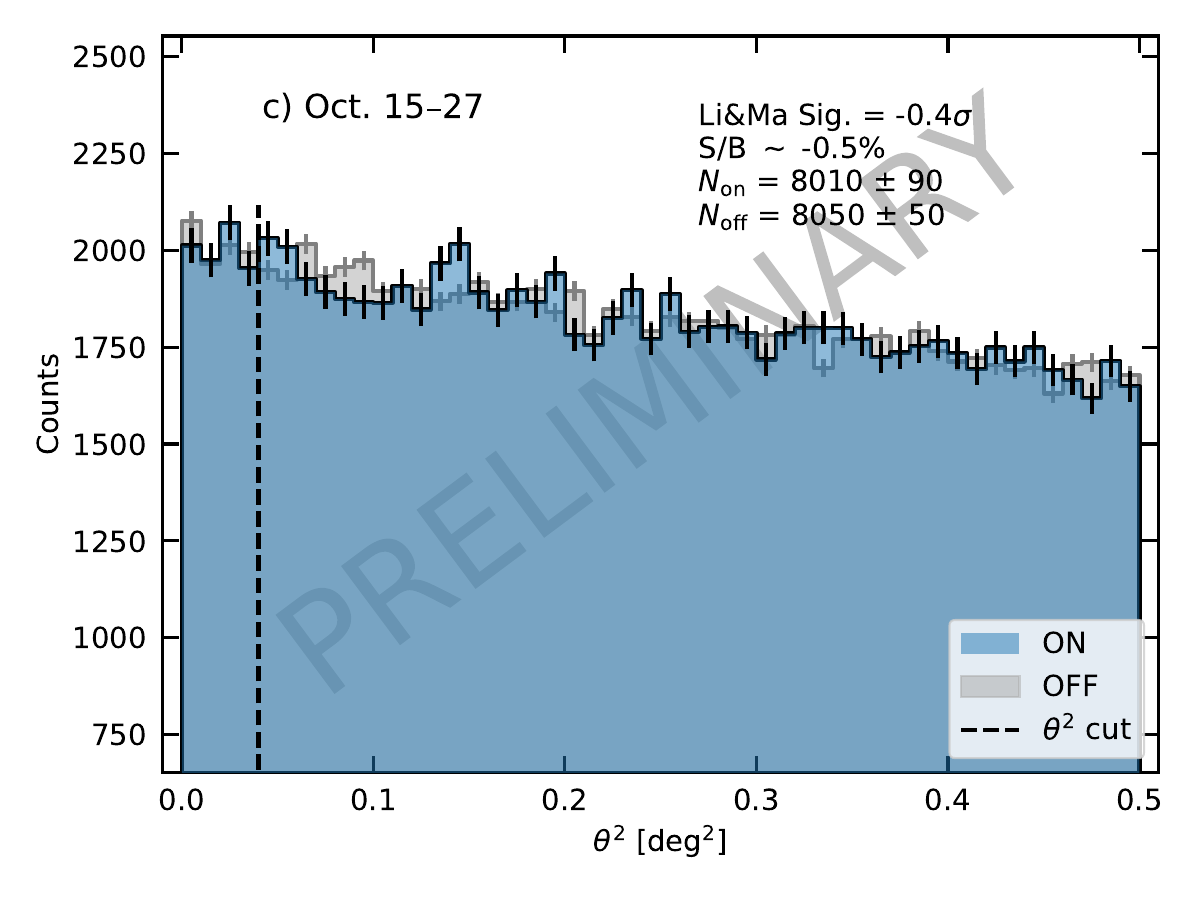}}
\caption{
\footnotesize
$\theta^2$ plots. From top to bottom, the angular distribution of events centred on GRB~221009A (ON) is compared with the mean angular distribution of events obtained at three OFF regions (OFF) using data on Oct.~10, Oct.~12 and Oct.\,15--27. The detection statistical significance (Li\&Ma Sig.) and signal-to-background ratio (S/B) is shown for events within the $\theta^2$ cut. Vertical errors bars are $1\sigma$ statistical uncertainties.}
\label{fig:theta2}
\end{figure}

On the first day of observations (Oct.~10, $T_0+$1.33\,d), the analysis yields a signal with a detection statistical significance of approximately $4\sigma$. This signal is confirmed with the independent analysis chain, providing a consistent result at the $4\sigma$ level. Subsequent observations on Oct.~12 and later ones (using non-consecutive days between Oct.~15--27; see Table~\ref{tab:obs_table}) show emission statistically compatible with the background fluctuations (see Fig.~\ref{fig:theta2}).

\begin{figure}[h!]
\resizebox{\hsize}{!}{\includegraphics[clip=true]{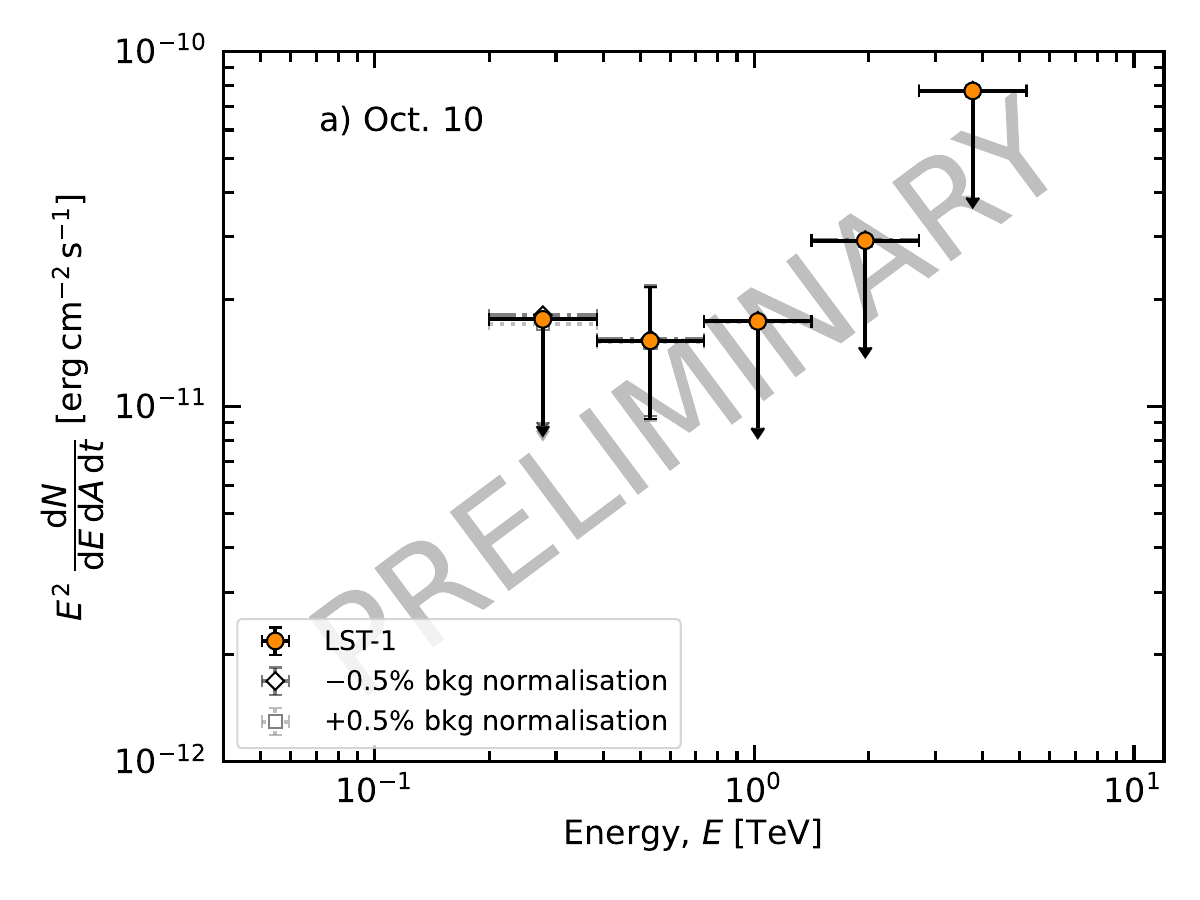}}\\
\resizebox{\hsize}{!}{\includegraphics[clip=true]{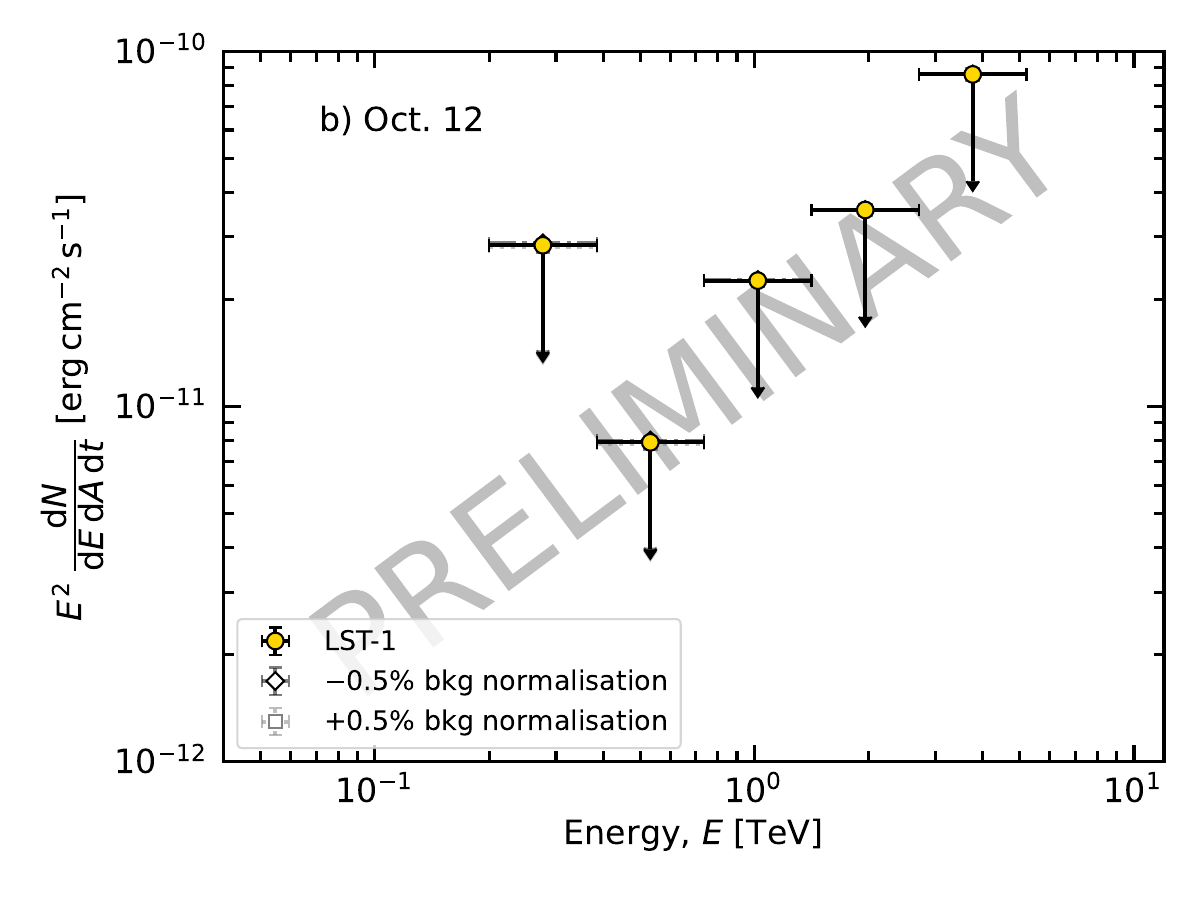}}\\
\resizebox{\hsize}{!}{\includegraphics[clip=true]{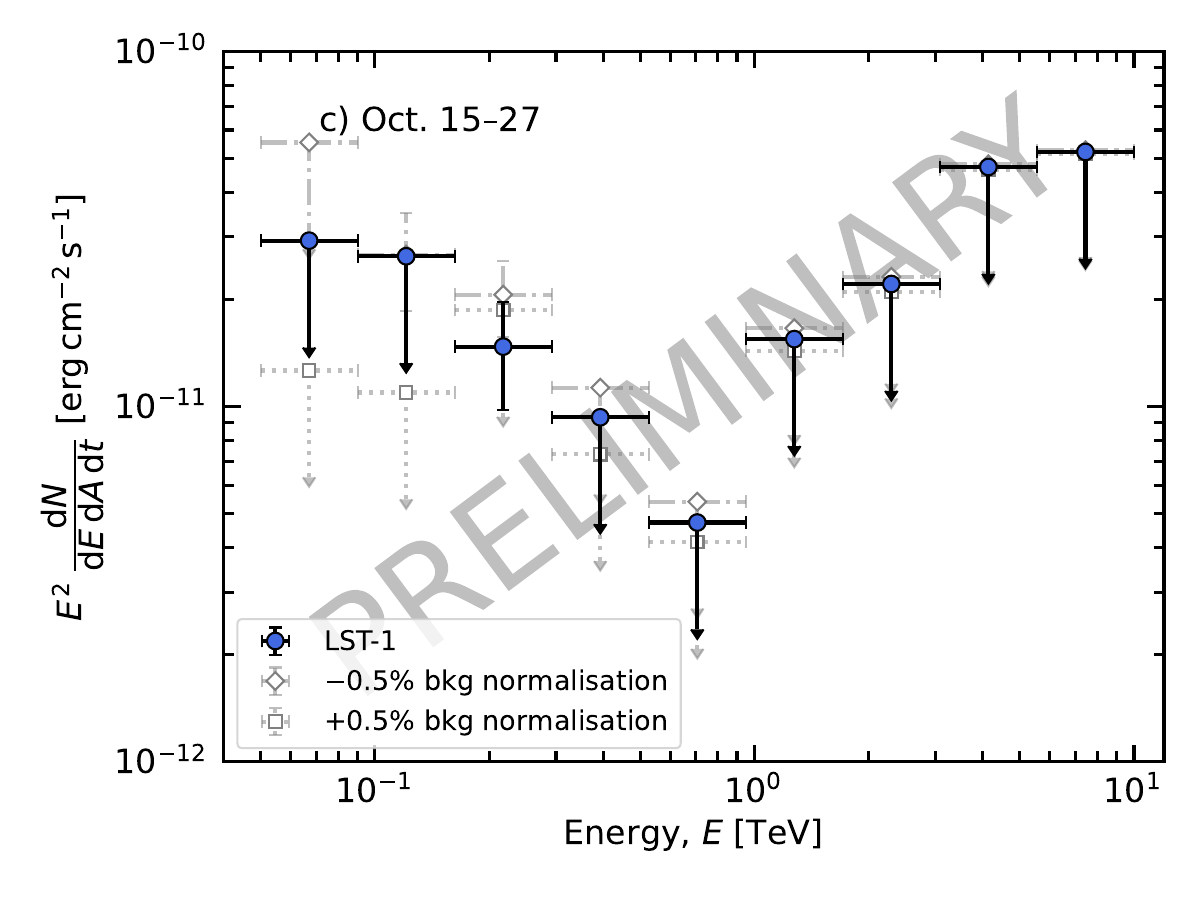}}
\caption{
\footnotesize
SEDs corrected for EBL attenuation. From top to bottom, the constraints on the SEDs using data on Oct.\,10, Oct.\,12 and Oct.\,15--27 are shown. For each plot, the tests of increasing and reducing by $0.5\%$ the background normalisation are shown using dotted-dashed and dotted error bars, respectively. This effect is negligible for the analysis of data on Oct.\,10 and Oct.\,12. Vertical errors bars are $1\sigma$ statistical uncertainties. Upper limits are computed at 95\% confidence level.}
\label{fig:sed}
\end{figure}

We constrain the SED of GRB~221009A on three time periods: Oct.~10, Oct.~12 and in non-consecutive observation days between Oct.~15--27. We use a power-law spectral model with an intrinsic spectral index fixed at $\Gamma=-2$ to describe the emission of GRB~221009A. The attenuation of the gamma-ray emission caused by the extragalactic background light (EBL) is considered using \citet{Dominguez2011} model with the estimated redshift of $z=0.1505$ \citep{GTCGCN2022}. In Fig.~\ref{fig:sed}, we show the SEDs corrected for EBL attenuation in the three periods. We obtain deep upper limits at around $10^{-11}$\,erg\,cm$^{-2}$\,s$^{-1}$ on an energy range poorly studied for this GRB at late times. 

The presence of high NSB observing conditions impacts the energy range that we can probe with LST-1, in particular increasing the energy threshold of the analysis. For the data on Oct.~10 and Oct.~12, we can constrain the emission down to 200\,GeV (see top and middle panels in Fig.~\ref{fig:sed}), whereas using data within Oct.~15--27, we probe the emission down to 50\,GeV, reflecting the LST-1 optimal capabilities to study the lowest energies under nominal observing conditions.

We assess the impact of systematic uncertainties on the background estimation for the spectral results. For that purpose, we tested a $\pm0.5\%$ systematic uncertainty in the background normalisation (the observed variation between events in the three control OFF regions). The outcome of these tests is shown as grey SED data points in Fig.~\ref{fig:sed}. We disfavour this source of systematics for the case of the moon-adapted analysis (Oct.~10 and Oct.~12 data) since the relative differences in the SEDs are smaller than $\sim$3\% (see top and middle panels in Fig.~\ref{fig:sed}). However, changes are observed at low energies for the SED obtained in nominal observing conditions between Oct.~15--27 (see bottom panel in Fig.~\ref{fig:sed}). Therefore, the significant SED point in panel c) should be considered with caution, as it may be enhanced by systematic uncertainties in the background normalisation. As pointed out in \citet{Abe_2023_LSTPerformance}, the monoscopic configuration of the GRB~221009A observations with LST-1 leads to a modest background suppression of dim events with energies close to the threshold of the telescope. Therefore, many background events in the low-energy bins survive the background subtraction, leading to a low signal-to-background ratio at these energies. As a result, a significant portion of these excess events could be attributed to background fluctuations if one assumes a $\pm0.5\%$ systematic uncertainty in the background normalisation.

\section{Conclusions}
\label{sect:conc}
An extensive follow-up campaign on GRB~221009A was performed with LST-1, the deepest performed by this instrument on a GRB. Beginning observations after $T_0+1.33$\,d, the first day constitutes the earliest follow-up by an IACT on this event. A hint of detection at $4\sigma$ is obtained on that day, while the signal is compatible with the background afterwards. Careful data processing and analysis were carried out to suppress the high NSB observing conditions that affected the data taking on the first two observation days. We constrain the emission of GRB 221009A down to a few tens/hundreds of GeV at a level of about $10^{-11}$\,erg\,cm$^{-2}$\,s$^{-1}$ at the time of the observations, an energy range previously unconstrained. The dedicated analysis of these data and the final results are provided in  \citet{2025ApJ...988L..42A}.

\vspace{0.3cm}
\noindent {\bf{Affiliations}}\par
$^{6}$Department of Physics, Tokai University, 4-1-1, Kita-Kaname, Hiratsuka, Kanagawa 259-1292, Japan\par

\begin{acknowledgements}
We gratefully acknowledge financial support from the agencies and organizations listed here: \href{https://www.lst1.iac.es/acknowledgements.html}{https://www.lst1.iac.es/acknowledgements.html}.
\end{acknowledgements}

\bibliographystyle{aa}
\bibliography{bibliography}

\end{document}